\documentclass[draft]{birkjour}
\usepackage[noadjust]{cite}
\usepackage{xcolor}
\usepackage{amssymb}
\RequirePackage[all]{xy}

\newtheorem{thm}{Theorem}[section]

\newtheorem{lem}[thm]{Lemma}

\theoremstyle{definition}

\theoremstyle{remark}

\numberwithin{equation}{section}

\newcommand{\BibTeX}{B\kern-0.1emi\kern-0.017emb\kern-0.15em\TeX}
\newcommand{\XYpic}{$\mathrm{X\kern-0.3em\raisebox{-0.18em}{Y}}$-$\mathrm{pic}\,$}

\newcommand{\cl}{C \kern -0.1em \ell}  %%Clifford algebra

\newcommand{\ed}{\end{document}}

\def\diag{{\rm diag}}

\def\First{{\rm First}}
\def\Last{{\rm Last}}

\def\R{{\mathbb R}}
\def\C{{\mathbb C}}
\def\F{{\mathbb F}}

\def\Adj{{\rm Adj}}
\def\mod{{\rm mod}}
\def\G{{\rm G}}
\def\I{{\rm I}}
\def\cl{{C}\!\ell}

\begin{document}

%-------------------------------------------------------------------------
% editorial commands: to be inserted by the editorial office
%
%\firstpage{1} \volume{228} \Copyrightyear{2004} \DOI{003-0001}
%\seriesextraline{This is the Concrete Title of this Book\br H.E. R and S.T.C. W, Eds.}
%\firstpage{1}

\title[Development of the Method of Averaging in Geometric Algebras]
 {Development of the Method of Averaging in Clifford Geometric Algebras}

\author[D. Shirokov]{Dmitry Shirokov}
\address{%
HSE University\\
Moscow 101000\\
Russia
\medskip}
\address{
and
\medskip}
\address{
Institute for Information Transmission Problems of the Russian Academy of Sciences \\
Moscow 127051 \\
Russia}
\email{dm.shirokov@gmail.com}

%\thanks{The publication was prepared within the framework of the Academic Fund Program at HSE University in 2022 (grant 22-00-001).}

\subjclass{15A66, 11E88}
\keywords{Clifford algebra; geometric algebra; method of averaging; Reynolds operator; Pauli's theorem}
%
%\date{\today}
%\dedicatory{Last Revised:\\ \today}
%%% ----------------------------------------------------------------------
\begin{abstract}
We develop the method of averaging in Clifford (geometric) algebras suggested by the author in previous papers. We consider operators constructed using two different sets of anticommuting elements of real or complexified Clifford algebras. These operators generalize Reynolds operators from the representation theory of finite groups. We prove a number of new properties of these operators. Using the generalized Reynolds operators, we give a complete proof of the generalization of Pauli's theorem to the case of Clifford algebras of arbitrary dimension. The results can be used in geometry, physics, engineering, computer science, and other applications.
\end{abstract}
\label{page:firstblob}
%%% ----------------------------------------------------------------------
\maketitle
%%% ----------------------------------------------------------------------
% \tableofcontents

\section{Introduction}\label{sect1}

In the present paper, we develop the method of averaging in Clifford (geometric) algebras suggested by the author in \cite{aver, averaging2, spin}. Namely, we introduce generalized Reynolds operators in Clifford algebras and prove a number of new properties of these operators. Theorems \ref{theoremObSv1}--\ref{theoremObSvDual} are new. We use generalized Reynolds operators to prove generalized Pauli's theorem (see Theorems \ref{theoremPauliEv} and \ref{theoremPauliOddReal}), which has been formulated for the first time in a brief report \cite{1} without a proof. The main idea of these theorems is to present an algorithm to compute the element $T$ that connects two sets of Clifford algebra elements that satisfy the main anticommutative conditions. The proofs of Theorems \ref{theoremPauliEv} and \ref{theoremPauliOddReal} are presented in this paper for the first time. The results are used in geometry, physics, and engineering, in particular, for $n$-dimensional Weyl, Majorana, and Majorana--Weyl spinors, to study relations between spin  and orthogonal groups, for the Dirac and the Yang--Mills equations in pseudo-Euclidean space and on non-trivial curved manifolds, etc.

This paper is organized as follows. In Section \ref{sect1}, we discuss Reynolds operators and generalized Reynolds operators in Clifford algebras and representation theory of finite groups. In Section \ref{sect2}, we prove some auxiliary statements about sets of anticommuting elements of the Clifford algebra.  In Section~\ref{sect3}, we prove a number of new properties of generalized Reynolds operators in Clifford algebras. In Section \ref{sect4}, we consider some other operators and prove their properties. In Section~\ref{sect5}, we prove that these operators are also related to generalized Reynolds operators in the case of odd dimension. In Section~\ref{sect6}, we use these operators to deduce an algorithm to compute an element that connects two sets of anticommuting elements of the Clifford algebra. The discussion follows in Section~\ref{sect7}.

Let $e_a$ be generators of the real Clifford algebra $\cl_{p,q}$ (or geometric algebra, see, for example, \cite{Lounesto, Hestenes2, BT,LM,DL}) and $e_A= e_{a_1 a_2\ldots a_k} = e_{a_1} e_{a_2} \cdots  e_{a_k}$ are basis elements enumerated by ordered multi-indices $A = a_1 a_2 \ldots a_k$, $a_1 < a_2 < \cdots < a_k$ with a length between $0$ and $n$ (the element $e:=e_{\emptyset}$ with empty multi-index is the identity element). The indices $a, a_1, a_2,\ldots$ take the values from $1$ to $n$. The generators satisfy the main anticommutative conditions of Clifford algebra 
$$e_a e_b+e_b e_a=2\eta_{ab}e,$$
where $\eta=||\eta_{ab}||=\diag(1,\ldots,1,-1,\ldots,-1)$ is the diagonal matrix with its first $p$ entries equal to $1$ and the last $q$ entries equal to $-1$ on the diagonal, $p+q=n$. We denote inverses of generators by $$e^a:=\eta^{ab}e_b=(e_a)^{-1},\qquad a=1, \ldots, n,$$
and inverses of basis elements by $e^A:=(e_A)^{-1}$. We use the Einstein summation convention.

Let us consider the Reynolds operator acting on the Clifford algebra
$$
\frac{1}{|G|}\sum_{g\in G}g^{-1}U g,\qquad U\in\cl_{p,q},
$$
where $|G|$ is the number of elements in a finite subgroup $G$ of the group $\cl^\times_{p,q}$ of all invertible elements of $\cl_{p,q}$. Reynolds operators are widely used in the representation theory of finite groups (see \cite{Babai, CLS, Dixon, Serr}).

In \cite{aver}, we consider the operator
\begin{eqnarray}
\frac{1}{2^n}\sum_{A\in\I} e_A U e^A,\label{RR1}
\end{eqnarray}
which is the Reynolds operator of the Salingaros vee group \cite{Sal1, Sal2, Sal3, Abl1, Abl2}
\begin{eqnarray}
&&\G_{p,q}=\{\pm e_A, A\in\I\},\qquad \dim \G_{p,q}=2^{n+1},\\
&&\I=\{\emptyset, \, 1,\, \ldots,\, n,\, 12,\, 13,\, \ldots,\, 1\ldots n\}\label{II}.
\end{eqnarray}
The operator (\ref{RR1}) is a projection onto the center ${\rm Cen}(\cl_{p,q})$ of the Clifford algebra $\cl_{p,q}$.

We can consider Reynolds operators acting on $\cl_{p,q}$ of the Salingaros vee group $\G_{p', q'}$ of another Clifford algebra $\cl_{p',q'}\subset\cl_{p,q}$. These operators ``average'' an action of the Salingaros vee group $\G_{p',q'}$ on the Clifford algebra $\cl_{p,q}$. For example, the operator
\begin{eqnarray}
\frac{1}{2}(U+(e_A)^{-1}U e_A)\label{RR2}
\end{eqnarray}
is the Reynolds operator of the group $\{\pm e, \pm e_A\}$ of order $4$, where $(e_A)^2=e$ or $(e_A)^2=-e$, which is isomorphic to the Salingaros vee group $\G_{1,0}$ (or $\G_{0,1}$). This operator is a projection onto the subset of elements that commute with the generator $e_A$ (see \cite{aver}, Section 6).

More generally, we can consider the operators
\begin{eqnarray}
\frac{1}{|S|}\sum_{A\in S} e_A U e^A,\qquad U\in\cl_{p,q},\label{Reyn0}
\end{eqnarray}
where $|S|$ is the number of elements of some subset $S$ of the set $\I$ (\ref{II}). In particular cases, we obtain (\ref{RR1}) and (\ref{RR2}). For some subsets $S$, the operators (\ref{Reyn0}) are not Reynolds operators. In the present paper, we consider the following new operators, which generalize (\ref{Reyn0}):
\begin{eqnarray}
\frac{1}{|S|}\sum_{A\in S}\beta_A U \gamma^A,\qquad U\in\cl_{p,q},\qquad \gamma^A:=(\gamma_A)^{-1},\label{Reyn00}
\end{eqnarray}
where the elements $\gamma_A$ and $\beta_A$ are generated by two different sets of Clifford algebra elements $\{\gamma_a, a=1, \ldots, n\}$ and $\{\beta_a, a=1, \ldots, n\}$ that satisfy the main anticommutative conditions of the Clifford algebras $\cl_{p,q}$. If $S=\I$, then the corresponding sets of elements $\{\pm \gamma_A, A\in\I\}$ and $\{\pm \beta_A, A\in \I\}$ form Salingaros vee groups of dimensions $2^{n+1}$ or $2^{n}$ (see Lemma \ref{theoremSecBasReal}). We call the operators (\ref{Reyn00}) {\it generalized Reynolds operators} in the case $S=\I$. 
Sometimes we consider operators (\ref{Reyn00}) with subsets $S$ such that the corresponding sets of elements $\{\pm \gamma_A, A\in S\}$ and $\{\pm \beta_A, A\in S\}$ do not form groups (see Section~\ref{sect4}), but are related to generalized Reynolds operators in the case of odd $n$ (see Section \ref{sect5}).
%In Section~\ref{sect3}, we prove a number of new properties of these operators. Sometimes we consider operators (\ref{Reyn00}) with subsets $S$ such that the corresponding sets of elements $\{\pm \gamma_A, A\in S\}$ and $\{\pm \beta_A, A\in S\}$ do not form groups (see Section~\ref{sect4}). In Section~\ref{sect5}, we prove that these operators are also related to generalized Reynolds operators in the case of odd $n$. In Section~\ref{sect6}, we use these operators to deduce an algorithm to compute an element $T$ that connects two sets $\{\gamma_a, a=1, \ldots, n\}$ and $\{\beta_a, a=1, \ldots, n\}$. The conclusions follow in Section~\ref{sect7}.

We formulate statements of this paper not only for the case of the real Clifford algebras $\cl_{p,q}$, but also for the case of the complexified Clifford algebras $\C\otimes\cl_{p,q}$ (see, for example, \cite{Lounesto} p. 139, \cite{BT} p. 80), which are important for applications. It is convenient for us to use the notation $\cl^\F_{p,q}$ ($\F=\R$ or $\C$), when results are true for both cases $\cl^\R_{p,q}:=\cl_{p,q}$, $\cl^\C_{p,q}:=\C\otimes\cl_{p,q}$.

\section{On sets of anticommuting Clifford algebra elements}\label{sect2}

Let us consider a set of Clifford algebra elements
\begin{eqnarray}
\{\gamma_a,\quad a=1, \ldots, n\}\in\cl^\F_{p,q},\qquad \gamma_a \gamma_b + \gamma_b \gamma_a= 2 \eta_{ab} e\label{oprs}
\end{eqnarray}
and the corresponding set
\begin{eqnarray}
\mathfrak{B}=\{\gamma_A, A\in\I\}=\{e, \gamma_a, \gamma_{ab}, \gamma_{abc}, \ldots, \gamma_{1\ldots n}\},\qquad a<b<c<\cdots ,\label{beta2}
\end{eqnarray}
where $\gamma_{a_1} \cdots \gamma_{a_k}$ is denoted by $\gamma_{a_1\ldots a_k}$ for $a_1<\cdots<a_k$.
We have the following lemma (see also \cite{Snygg}, pp. 289--290, and \cite{Port}, pp. 127--128). We present the proof for the convenience of the reader.
\begin{lem}\label{theoremSecBasReal}
\begin{enumerate}
  \item If $n=p+q$ is even, then the set (\ref{beta2}) is linearly independent.
  \item If $n=p+q$ is odd, then either
   \begin{itemize}
     \item we have $\gamma_{1\ldots n}=\pm e_{1\ldots n}$ and (\ref{beta2}) is linearly independent, or
     \item we have $\gamma_{1\ldots n}=\pm e$ and (\ref{beta2}) is linearly dependent (this is possible only in the case $p-q\equiv 1 \mod4$), or
     \item we have $\gamma_{1\ldots n}=\pm ie$ and (\ref{beta2}) is linearly dependent (this is possible only in case $\F=\C$, $p-q\equiv 3 \mod4$).
   \end{itemize}
\end{enumerate}
\end{lem}

Note that in the cases $\F=\R$ when the set (\ref{beta2}) is linearly independent, this set is another basis of $\cl_{p,q}$. In the cases when the set is linearly dependent, it is the set of two bases of the subalgebra of $\cl_{p,q}$, which is isomorphic to another Clifford algebra $\cl_{p', q'}$ of dimension $2^{n-1}$, $p'+q'=n-1$. For example, in $\cl_{2,1}$ with the set of generators $e_1, e_2, e_3$ we can take $\gamma_1=e_1$, $\gamma_2=e_2$, $\gamma_3=e_{12}$ and obtain the set $\mathfrak{B}=\{e, e_1, e_2, e_{12}, e_{12}, e_2, -e_1, -e\}$
of two bases of $\cl_{2,0}$.

\begin{proof} Suppose that the set (\ref{beta2}) is linearly dependent. Then there exist not all zero scalars $u, u_1, \ldots, u_{1\ldots n}$ such that
$$ue+u_1\gamma_1+\cdots+u_{1\ldots n}\gamma_{1\ldots n}=0.$$
At least one scalar is non-zero $u_B\neq 0$ for some multi-index $B$. Multiplying both sides of the equation by $\frac{\gamma_B}{u_B}$, we get
\begin{eqnarray}
e+v_1\gamma_1+\cdots+v_{1\ldots n}\gamma_{1\ldots n}=0\label{tt1}
\end{eqnarray}
for some other scalars $v_1, \ldots, v_{1\ldots n}$. At least one of these scalars is nonzero because $e\neq 0$.

For any element $\gamma_A$ (except $\gamma_{1\ldots n}$ in the case of odd $n$ and except $e$ in the case of any $n$), there exists element $\gamma_a$ such that $\gamma_A$ anticommutes with $\gamma_a$ (if $|A|$ is even, then we can take $a\in A$; if $|A|$ is odd, then we can take $a\notin A$). Let us choose some $\gamma_A$ and some $\gamma_a$ that anticommutes with $\gamma_A$. Multiplying both sides of (\ref{tt1}) on the right by $\gamma_a$ and on the left by $(\gamma_a)^{-1}$, we obtain the equation like (\ref{tt1}) but with another sign of the term $v_A \gamma_A$ (the signs of some other terms may also change). Adding both sides of this new equation and both sides of (\ref{tt1}), we obtain again the equation like (\ref{tt1}) but without the term $v_A \gamma_A$ (and some other terms) and with other scalars. Further, we repeat this process for other terms.

In the case of even $n$, after finite number of steps, we obtain $e=0$, i.e. a contradiction. Hence, the set (\ref{beta2}) is linearly independent.

In the case of odd $n$, after finite number of steps, we get $e+w_{1\ldots n}\gamma_{1\ldots n} = 0$ for some scalar $w_{1\ldots n}$, i.e. $\gamma_{1\ldots n}=\lambda e$ for some scalar $\lambda$. We have
\begin{eqnarray}
(\gamma_{1\ldots n})^2=(-1)^{\frac{n(n-1)}{2}}(-1)^q e.\label{uu1}
\end{eqnarray}
If $\frac{n(n-1)}{2}+q$ is even, which is equivalent to $p-q\equiv 1\mod 4$, then $\gamma_{1\ldots n}=\pm e$. If $\frac{n(n-1)}{2}+q$ is odd, which is equivalent to $p-q\equiv 3\mod 4$, then $\gamma_{1\ldots n}=\pm i e$, which is possible only in the complex case. In other cases, the set (\ref{beta2}) is linearly independent. Since the conditions (\ref{oprs}), it follows that  $\gamma_{1\ldots n}$ is in the center\footnote{Note that the center ${\rm Cen}(\cl^\F_{p,q})=\{U\in \cl^\F_{p,q} \,|\, UV=VU, \forall V\in\cl^\F_{p,q}\}$ of $\cl^\F_{p,q}$ is the subspace $\cl^{ 0\, \F}_{p,q}$ in the case of even $n$ and the subspace $\cl^{0\,\F}_{p,q}\oplus\cl^{n\,\F}_{p,q}$ in the case of odd $n$.} of the Clifford algebra. We have $\gamma_{1\ldots n}=ae+be_{1\ldots n}$ for some scalars $a$ and $b$. We get
\begin{eqnarray}
(\gamma_{1\ldots n})^2=(a^2+b^2 (-1)^{\frac{n(n-1)}{2}+q})e+2ab e_{1\ldots n}.\label{uu2}
\end{eqnarray}
From (\ref{uu1}) and (\ref{uu2}), we obtain $ab=0$. If $b=0$, then $\gamma_{1\ldots n}=ae$, and we obtain a contradiction, because the set (\ref{beta2}) is linearly independent. Hence, $a=0$ and $b^2=1$, i.e. $\gamma_{1\ldots n}=\pm e_{1\ldots n}$.
\end{proof}

Let us consider the operator
\begin{eqnarray}F(U)=\frac{1}{2^n} \sum_{A\in\I}\gamma_A U \gamma^A,\qquad \gamma^A:=(\gamma_A)^{-1},\qquad U\in\cl^\F_{p,q},\nonumber\end{eqnarray}
where we have a sum over multi-index $A\in\I$ (\ref{II}). Note that in the real case $\F=\R$, if $\{\gamma_A, A\in\I\}$ is a basis of $\cl_{p,q}$, then $F$ is the Reynolds operator of the Salingaros vee group, see \cite{aver}.

Let us denote a vector subspace spanned by the elements $e_{a_1\ldots a_k}$ enumerated by the ordered multi-indices of length $k$ by $\cl^{k\, \F}_{p,q}$. Any Clifford algebra element $U\in\cl^\F_{p,q}$ can be written in the form
\begin{eqnarray*}
U=ue+\!\sum_a\! u_a e_a+\!\!\sum_{a_1<a_2}\!\!u_{a_1 a_2}e_{a_1 a_2}+\cdots+u_{1\ldots n}e_{1\ldots n},\,\,\,\, u, u_a, \ldots, u_{1\ldots n}\in \F.
\end{eqnarray*}
We use the projection operator\footnote{Note that this operation coincides up to a constant with the trace of the corresponding matrix representation, see \cite{trace,det}.} $\pi_0(U)=u$ onto the vector subspace $\cl^{0\, \F}_{p,q}$ (where $e\equiv 1$). Also we consider the following operator: $\pi_n(U)=u_{1\ldots n}$ for $U\in\cl^\F_{p,q}$.

If the set $\{\gamma_A, A\in\I\}$ is linearly independent, then we have (this is proved for the real case $\F=\R$ in \cite{aver})
\begin{eqnarray}F(U)=\frac{1}{2^n}\sum_{A}\gamma_A U \gamma^A=\left\lbrace
\begin{array}{ll}
\pi_0(U)e, & \parbox{.2\linewidth}{if $n$ is even;} \\
\pi_0(U)e+\pi_n(U)e_{1\ldots n}, & \parbox{.2\linewidth}{if $n$ is odd.}
\end{array}
\right.\label{poln}
\end{eqnarray}
Operator $F$ is a projection $F^2=F$ onto the center of $\cl^\F_{p,q}$.
Note that here and after we use notation $F^2(U):=F(F(U))\neq (F(U))^2$ and similar notation for the squares of all other operators in this paper.
In the language of \cite{CLS}, ${\rm Cen}(\cl_{p,q})$ is the ``ring of invariants'' of Salingaros vee group.

\section{Generalized Reynolds operators}\label{sect3}

Let us consider two different sets of Clifford algebra elements
\begin{eqnarray}
&&\{\gamma_a,\quad a=1, \ldots, n\}\in\cl^\F_{p,q},\qquad \gamma_a \gamma_b + \gamma_b \gamma_a= 2 \eta_{ab} e, \label{gamma}\\
&&\{\beta_a,\quad a=1, \ldots, n\}\in\cl^\F_{p,q},\qquad \beta_a \beta_b + \beta_b \beta_a= 2 \eta_{ab} e,\label{beta}
\end{eqnarray}
and {\it generalized Reynolds operators}
\begin{eqnarray}
H(U)=\frac{1}{2^n}\sum_{A\in\I}\beta_A U \gamma^A,\quad P(V)=\frac{1}{2^n}\sum_{A\in\I} \gamma_A V \beta^A,\qquad U, V\in\cl^\F_{p,q},\label{Q}
\end{eqnarray}
where we have sum over multi-index $A\in\I\!=\!\{\emptyset, \, 1,\, \ldots,\, n,\, 12,\, 13,\, \ldots,\, 1\ldots n\}$ of $2^n$ terms, $\beta_A=\beta_{a_1\ldots a_k}:=\beta_{a_1} \cdots \beta_{a_k}$ for $a_1<\cdots<a_k$, $\gamma^A:=(\gamma_A)^{-1}$, $\beta^A:=(\beta_A)^{-1}$. We consider both operators (\ref{Q}) at the same time, because we want to study how they are related (see Theorem \ref{theoremObSv3}).

\begin{thm}\label{theoremObSv1} We have
\begin{eqnarray}
\beta_B H(U)=H(U) \gamma_B,\qquad \gamma_B P(V)=P(V) \beta_B,\qquad \forall B\in \I.\label{obshcommut}
\end{eqnarray}
The operators $H$ and $P$ are projections:
$$H^2=H,\qquad P^2=P,$$
where $H^2(U):=H(H(U))$ and $P^2(U):=P(P(U))$.
\end{thm}

\begin{proof} We have
\begin{eqnarray}
\beta_B H(U) \gamma^B&=&\beta_B \frac{1}{2^n}\sum_A \beta_A U \gamma^A \gamma^B=\frac{1}{2^n}\sum_A (\beta_B \beta_A) U (\gamma_B \gamma^A)\nonumber\\
&=&\frac{1}{2^n}\sum_A \beta_A U \gamma^A=H(U).\nonumber
\end{eqnarray}
Thus,
$$H^2(U)=\frac{1}{2^n}\sum_B \beta_B H(U) \gamma^B=\frac{1}{2^n} (2^n H(U))=H(U).$$
Proof of the statement for the operator $P$ is similar.  \end{proof}

\begin{thm}\label{theoremObSv3} Let us consider $\cl^\F_{p,q}$, linearly independent sets $\{\gamma_A, A\in\I\}$, $\{\beta_A, A\in\I\}$, and the operators (\ref{Q}).
\begin{enumerate}
                \item If $n=p+q$ is even, then
                \begin{eqnarray}
                P(V)H(U)=H(U)P(V)=\pi_0(V H(U)) e,\label{QT}
                \end{eqnarray}
                where $$\pi_0(V H(U))=\pi_0(H(U) V)=\pi_0(U P(V))=\pi_0(P(V) U).$$
                \item If $n=p+q$ is odd, then
                \begin{eqnarray}
                 P(V)H(U)=H(U)P(V)=\pi_0(V H(U)) e+\pi_n(V H(U)) e_{1\ldots n},\label{QTnech}
                \end{eqnarray}
                where $$\pi_0(V H(U))=\pi_0(H(U) V)=\pi_0(U P(V))=\pi_0(P(V) U),$$
                $$\pi_n(V H(U))=\pi_n(H(U) V)=\pi_n(U P(V))=\pi_n(P(V) U).$$
              \end{enumerate}
\end{thm}

\begin{proof} Using (\ref{poln}) and Theorem \ref{theoremObSv1}, we obtain
\begin{eqnarray}
P(V)H(U)&=&\frac{1}{2^n}\sum_A \gamma_A V \beta^A H(U)=\frac{1}{2^n}\sum_A \gamma_A V H(U) \gamma^A\nonumber\\
&=&\left\lbrace
\begin{array}{ll}
\pi_0(V H(U)) e, & \parbox{.2\linewidth}{if $n$ is even;} \\
\pi_0(V H(U)) e+ \pi_n(V H(U)) e_{1\ldots n}, & \parbox{.2\linewidth}{if $n$ is odd.}
\end{array}
\right.\nonumber
\end{eqnarray}

We can similarly obtain
\begin{eqnarray}
P(V)H(U)&=&\frac{1}{2^n}P(V) \sum_A\beta_A U \gamma^A=\frac{1}{2^n}\sum_A\gamma_A P(V) U \gamma^A\nonumber\\
&=&\left\lbrace
\begin{array}{ll}
\pi_0(P(V)U) e, & \parbox{.2\linewidth}{if $n$ is even;} \\
\pi_0(P(V)U) e+ \pi_n(P(V)U) e_{1\ldots n}, & \parbox{.2\linewidth}{if $n$ is odd.}
\end{array}
\right.\nonumber
\end{eqnarray}
Further we use the properties $\pi_0(AB)=\pi_0(BA)$ for arbitrary $A, B\in\cl^\F_{p,q}$ (see \cite{trace,det}) and $\pi_n(AB)=\pi_n(BA)$ for arbitrary $A, B\in\cl^\F_{p,q}$ with odd $n=p+q$ (see \cite{Sh1}).  \end{proof}

We use the following notation
\begin{eqnarray}
\I_{(0)}=\{A\in\I,\, |A| \mbox{ is even}\},\qquad \I_{(1)}=\{A\in\I,\, |A| \mbox{ is odd}\}.\nonumber
\end{eqnarray}
Note that the parity of an element $\gamma_A$ is not the parity of length of the corresponding multi-index $A$. The same element $\gamma_A$ may have multi-indices with lengths of different parity in different representations.
%not be uniquely determined for each element $\gamma_A$ and depends on the representation (it can be even for one multi-index and odd for another multi-index).
In the case $\gamma_{1 \ldots n}=\pm e$, $n$ is odd (see Lemma \ref{theoremSecBasReal}), we have $2^{n-1}$ pairs of coincident (up to sign) elements in the set $\mathfrak{B}=\{ \gamma_A, A\in\I\}$. In this case, the same element $\gamma_{1 \ldots n}=\pm e$ has multi-index with even length in one representation ($\pm e$) and multi-index with odd length in another representation ($\gamma_{1\ldots n}$).

To prove Theorem \ref{theoremObSv5}, we need the following statement.

\begin{lem}\label{theoremCommBas2} Let us consider $\cl^\F_{p,q}$, $p+q=n$ and the set of elements $\mathfrak{B}=\{\gamma_A, A\in\I\}$ with the property (\ref{gamma}). Then each element of this set (if it is neither $e$ nor $\gamma_{1\ldots n}$) commutes with $2^{n-2}$ elements of the set $\mathfrak{B}$ with multi-index of even length, commutes with $2^{n-2}$ elements of the set $\mathfrak{B}$ with multi-index of odd length, anticommutes with $2^{n-2}$ elements of the set $\mathfrak{B}$ with multi-index of even length and anticommutes with $2^{n-2}$ elements of the set $\mathfrak{B}$ with multi-index of odd length.
The element $e$ commutes with all elements of the set $\mathfrak{B}$.
\begin{enumerate}
  \item if $n$ is even, then $\gamma_{1\ldots n}$ commutes with all $2^{n-1}$ elements of the set $\mathfrak{B}$ with multi-index of even length and anticommutes with all $2^{n-1}$ elements of the set $\mathfrak{B}$ with multi-index of odd length;
  \item if $n$ is odd, then $\gamma_{1\ldots n}$ commutes with all $2^n$ elements of the set $\mathfrak{B}$.
\end{enumerate}
\end{lem}
\begin{proof}
In \cite{aver}, we proved the particular case of this statement for the set $\{e_A, A\in\I\}$. In this proof, we did not use the fact that this set is linearly independent (we use only anticommutative properties of the elements). Hence, the statement is valid for the set $\mathfrak{B}=\{\gamma_A, A\in\I\}$, which is not always linearly independent (see Lemma \ref{theoremSecBasReal}).
\end{proof}

\begin{thm}\label{theoremObSv5} Let us consider $\cl^\F_{p,q}$, $n=p+q$ and two sets (\ref{gamma}), (\ref{beta}).
Then
 \begin{eqnarray}
\sum_{B\in\I_{(0)}} H(\gamma_B)\gamma^B&=&\frac{1}{2^n}\sum_A\sum_{B\in\I_{(0)}}\beta_A \gamma_B \gamma^A \gamma^B\\
&=&\left\lbrace
\begin{array}{ll}
\frac{1}{2}(e + \beta_{1\ldots n}\gamma^{1\ldots n}), & \parbox{.2\linewidth}{if $n$ is even;} \\
\frac{1}{2}(e + \beta_{1\ldots n}\gamma^{1\ldots n}), & \parbox{.2\linewidth}{if $n$ is odd,}
\end{array}
\right.\label{r2}
\end{eqnarray}
\begin{eqnarray}
\sum_{B\in\I_{(1)}} H(\gamma_B)\gamma^B&=&\frac{1}{2^n}\sum_A\sum_{B\in\I_{(1)}}\beta_A \gamma_B \gamma^A \gamma^B\\
&=&\left\lbrace
\begin{array}{ll}
\frac{1}{2}(e - \beta_{1\ldots n}\gamma^{1\ldots n}), & \parbox{.2\linewidth}{if $n$ is even;} \\
\frac{1}{2}(e + \beta_{1\ldots n}\gamma^{1\ldots n}), & \parbox{.2\linewidth}{if $n$ is odd,}
\end{array}
\right.\label{r3}
\end{eqnarray}
\begin{eqnarray}
\sum_B H(\gamma_B)\gamma^B&=&\frac{1}{2^n}\sum_A\sum_B\beta_A \gamma_B \gamma^A \gamma^B\\
&=&\left\lbrace
\begin{array}{ll}
e, & \parbox{.2\linewidth}{if $n$ is even;} \\
e + \beta_{1\ldots n}\gamma^{1\ldots n}, & \parbox{.2\linewidth}{if $n$ is odd.}
\end{array}
\right.\label{r23}
\end{eqnarray}
\end{thm}

\begin{proof} Let us consider the following expressions in the case of even $n$ (we swap all $\gamma_B$ and $\gamma^A$ and obtain plus or minus in each case):
\begin{eqnarray}
\sum_A \beta_A e\gamma^A&=&e+\beta_1 \gamma^1+\cdots+\beta_{1\ldots n}\gamma^{1\ldots n}\\
&=&(e+\beta_1 \gamma^1+\cdots+\beta_{1\ldots n}\gamma^{1\ldots n})e,\nonumber\\
\sum_A\beta_A \gamma_1 \gamma^A&=&\gamma_1+\beta_1 \gamma_1 \gamma^1+\cdots+\beta_{1\ldots n}\gamma_1 \gamma^{1\ldots n}\\
&=&(e+\beta_1 \gamma^1-\cdots-\beta_{1\ldots n}\gamma^{1\ldots n})\gamma_1,\nonumber\\
%\sum_A \beta_A \gamma_2 \gamma^A&=&\gamma_2+\beta_1 \gamma_2 \gamma^1+\cdots+\beta_{1\ldots n}\gamma_2 \gamma^{1\ldots n}\nonumber\\
%&=&(e-\beta_1 \gamma^1+\cdots-\beta_{1\ldots n} \gamma^{1\ldots n})\gamma_2,\nonumber\\
&\cdots&\nonumber\\
\sum_A\beta_A \gamma_{1\ldots n} \gamma^A&=&\gamma_{1\ldots n}+\beta_1 \gamma_{1\ldots n} \gamma^1+\cdots+\beta_{1\ldots n}\gamma_{1\ldots n} \gamma^{1\ldots n}\\
&=&(e-\beta_1 \gamma^1-\cdots+\beta_{1\ldots n}\gamma^{1\ldots n})\gamma_{1\ldots n}.\nonumber
\end{eqnarray}
Let us multiply (on the right) both sides of the first equation by $e$, the second equation by $\gamma^1$, \ldots, the last equation by $\gamma^{1\ldots n}$. Summing equalities with $B\in\I_{(0)}$ (or $B\in\I_{(1)}$) and using Lemma \ref{theoremCommBas2} (we must take into account the number of pluses and minuses), we obtain the statement of this theorem. In the case of odd $n$, the proof is similar. Summing (\ref{r2}) and (\ref{r3}), we obtain (\ref{r23}).  \end{proof}

\begin{thm}\label{theoremObSv6} Let us consider $\cl^\F_{p,q}$, $n=p+q$, two sets (\ref{gamma}), (\ref{beta}), and the operator $H(U)$ (\ref{Q}).
\begin{enumerate}
                \item If $n$ is even, then there exists an element $U\in\{\gamma_A, A\in\I\}$ such that $H(U)$ in nonzero.
                Moreover, we can take $U\in\{ \gamma_A, A\in\I_{(0)}\}$ if $\beta_{1\ldots n}\neq-\gamma_{1\ldots n}$ and we can take $U\in\{ \gamma_A, A\in\I_{(1)}\}$ if $\beta_{1\ldots n}\neq\gamma_{1\ldots n}$.
                \item If $n$ is odd and $\beta_{1\ldots n}\neq -\gamma_{1\ldots n}$, then there exists an element $U\in\{\gamma_A, A\in\I\}$ such that $H(U)$ is nonzero.
                Moreover, we can take $U\in\{ \gamma_A, A\in\I_{(0)}\}$ and we can take $U\in\{ \gamma_A, A\in\I_{(1)}\}$ at the same time.
 \end{enumerate}
 \end{thm}

\begin{proof} Let $n$ be even. Suppose that for all elements $U$ the operator $H(U)=\frac{1}{2^n}\sum_A\beta_A U \gamma^A$ equals zero. It follows from Theorem \ref{theoremObSv5} (see (\ref{r23})) that
$$2^n e=\sum_B(\sum_A\beta_A \gamma_B \gamma^A) \gamma^B=\sum_{B}0\,\gamma^B=0$$ and we obtain a contradiction.
Using (\ref{r2}), (\ref{r3}) we can similarly prove the statement in the other cases.  \end{proof}

We use Theorems \ref{theoremObSv1}, \ref{theoremObSv3}, and \ref{theoremObSv6} in Section \ref{sect6}.

\section{Some other operators}\label{sect4}

Let us consider the following operators for the sets (\ref{gamma}) and (\ref{beta})
\begin{eqnarray}
H'(U)=\frac{1}{2^n}(\!\!\!\sum_{A\in\I_{(0)}}\!\!\!\!\beta_A U \gamma^A\!-\!\!\!\! \sum_{A\in\I_{(1)}}\!\!\!\!\beta_A U \gamma^A)=\!\!\sum_A \frac{(-1)^{|A|}}{2^n}\beta_A U \gamma^A,\quad U\in\cl^\F_{p,q},\label{W}\\
P'(V)=\frac{1}{2^n}(\!\!\!\sum_{A\in\I_{(0)}}\!\!\!\!\gamma_A V \beta^A\!-\!\!\!\!\sum_{A\in\I_{(1)}}\!\!\!\!\gamma_A V \beta^A)=\!\!\sum_A\frac{(-1)^{|A|}}{2^n}\gamma_A V \beta^A,\quad V\in\cl^\F_{p,q}.\nonumber
\end{eqnarray}

\begin{thm}\label{theoremObSv7} We have
\begin{eqnarray}
&&\beta_B H'(U)=H'(U) \gamma_B,\qquad \gamma_B P'(V)=P'(V) \beta_B,\qquad  \forall B\in\I_{(0)},\nonumber\\
&&\beta_B H'(U)=-H'(U) \gamma_B,\qquad \gamma_B P'(V)=-P'(V) \gamma_B,\qquad \forall B\in\I_{(1)},\nonumber
\end{eqnarray}
in particular
\begin{eqnarray}
\beta_b H'(U)=-H'(U) \gamma_b,\qquad \gamma_b P'(V)=-P'(V) \beta_b,\qquad \forall b=1, \ldots, n.\nonumber
\end{eqnarray}
The operators $H'$ and $P'$ are projections:
$$H'(H'(U))=H'(U),\qquad P'(P'(V))=P'(V).$$
Also, we have
$$H(H'(U))=0,\qquad H'(H(U))=0,\qquad P(P'(V))=0,\qquad P'(P(V))=0.$$
\end{thm}

\begin{thm}\label{theoremObSv9} Let us consider $\cl^\F_{p,q}$ and linearly independent sets $\{\gamma_A, A\in\I\}$, $\{\beta_A, A\in\I\}$.
    \begin{enumerate}
                \item If $n=p+q$ is even, then
                \begin{eqnarray}
                 H'(U)P'(V)=P'(V)H'(U)&=&\pi_0(H'(U)V) e,\nonumber
                \end{eqnarray}
                where $$\pi_0(H'(U)V)=\pi_0(V H'(U))=\pi_0(P'(V) U)=\pi_0(U P'(V)).$$
                \item If $n=p+q$ is odd, then
                \begin{eqnarray}
                 H'(U)P'(V)=\pi_0(H'(U)V) e+\pi_n(H'(U)V) e_{1\ldots n},\nonumber
                \end{eqnarray}
                where $$\pi_0(H'(U)V)=\pi_0(V H'(U))=\pi_0(P'(V) U)=\pi_0(U P'(V)),$$
                $$\pi_n(H'(U)V)=\pi_n(V H'(U))=\pi_n(P'(V) U)=\pi_n(U P'(V)).$$
              \end{enumerate}
\end{thm}

\begin{thm}\label{theoremObSv10} Let us consider $\cl^\F_{p,q}$, $n=p+q$ and two sets (\ref{gamma}), (\ref{beta}). Then
\begin{eqnarray}
\sum_{B\in\I_{(0)}} H'(\gamma_B) \gamma^B&=& \frac{1}{2^n}\sum_A\sum_{B\in\I_{(0)}}(-1)^{|A|}\beta_A \gamma_B \gamma^A \gamma^B\\
&=&\left\lbrace
\begin{array}{ll}
\frac{1}{2}(e + \beta_{1\ldots n}\gamma^{1\ldots n}), & \parbox{.2\linewidth}{if $n$ is even;} \\
\frac{1}{2}(e - \beta_{1\ldots n}\gamma^{1\ldots n}), & \parbox{.2\linewidth}{if $n$ is odd,}
\end{array}
\right. \nonumber
\end{eqnarray}
\begin{eqnarray}
\sum_{B\in\I_{(1)}} H'(\gamma_B) \gamma^B&=&\frac{1}{2^n}\sum_A\sum_{B\in\I_{(1)}}(-1)^{|A|}\beta_A \gamma_B \gamma^A \gamma^B\\
&=&\left\lbrace
\begin{array}{ll}
\frac{1}{2}(e - \beta_{1\ldots n}\gamma^{1\ldots n}), & \parbox{.2\linewidth}{if $n$ is even;} \\
\frac{1}{2}(e - \beta_{1\ldots n}\gamma^{1\ldots n}), & \parbox{.2\linewidth}{if $n$ is odd,}
\end{array}
\right.\nonumber
\end{eqnarray}
\begin{eqnarray}
\sum_B H'(\gamma_B)\gamma^B&=&\sum_A \sum_B \frac{1}{2^n}(-1)^{|A|}\beta_A \gamma_B \gamma^A \gamma^B\\
&=&\left\lbrace
\begin{array}{ll}
e, & \parbox{.2\linewidth}{if $n$ is even;} \\
e - \beta_{1\ldots n}\gamma^{1\ldots n}, & \parbox{.2\linewidth}{if $n$ is odd.}
\end{array}
\right.\nonumber
\end{eqnarray}
\end{thm}

\begin{thm}\label{theoremObSv12} Let us consider $\cl^\F_{p,q}$, $n=p+q$, two sets (\ref{gamma}), (\ref{beta}),  and the operator $H'(U)$ (\ref{W}).
\begin{enumerate}
                \item If $n$ is even, then there exists an element $U\in\{\gamma_A, A\in\I\}$ such that $H'(U)$ is nonzero. Moreover, we can take $U\in\{\gamma_A, A\in \I_{(0)}\}$ if $\beta_{1\ldots n}\neq-\gamma_{1\ldots n}$ and we can take $U\in\{\gamma_A, A\in\I_{(1)}\}$ if $\beta_{1\ldots n}\neq\gamma_{1\ldots n}$.
                \item If $n$ is odd and $\beta_{1\ldots n}\neq \gamma_{1\ldots n}$, then there exists an element $U\in\{\gamma_A, A\in\I \}$ such that $H'(U)$ is nonzero. Moreover, we can take $U\in\{\gamma_A, A\in\I_{(0)}\}$ and we can take $U\in\{\gamma_A, A\in\I_{(1)}\}$ at the same time.
 \end{enumerate}
 \end{thm}

Note, that we have the condition $\beta_{1\ldots n}\neq \gamma_{1\ldots n}$ in Theorem \ref{theoremObSv12} and the condition $\beta_{1\ldots n}\neq -\gamma_{1\ldots n}$ in Theorem \ref{theoremObSv6} in the case of odd $n$.
\begin{proof}
We omit the proof of Theorems \ref{theoremObSv7}--\ref{theoremObSv12}, because it is similar to the proof of Theorems \ref{theoremObSv1}--\ref{theoremObSv6}.
\end{proof}

Let us consider the following operators for the sets (\ref{gamma}) and (\ref{beta})
\begin{eqnarray}
&&H_{(0)}(U)=\frac{1}{2^{n-1}}\!\!\!\sum_{A\in\I_{(0)}}\!\!\!\beta_A U \gamma^A,\quad H_{(1)}(U)=\frac{1}{2^{n-1}}\!\!\!\sum_{A\in\I_{(1)}}\!\!\!\beta_A U \gamma^A,\quad U\in\cl^\F_{p,q},\nonumber\\
&&P_{(0)}(V)=\frac{1}{2^{n-1}}\!\!\!\sum_{A\in\I_{(0)}}\!\!\!\gamma_A V \beta^A,\quad P_{(1)}(V)=\frac{1}{2^{n-1}}\!\!\!\sum_{A\in\I_{(1)}}\!\!\!\gamma_A V \beta^A,\quad  V\in\cl^\F_{p,q}.\nonumber
\end{eqnarray}
The corresponding operators in the particular case $\gamma_A=\beta_A=e_A$, $\F=\R$, are considered in \cite{averaging2}. We have
\begin{eqnarray}
&&H(U)=\frac{1}{2}(H_{(0)}(U)+H_{(1)}(U)),\qquad H'(U)=\frac{1}{2}(H_{(0)}(U)-H_{(1)}(U)),\nonumber\\
&&P(V)=\frac{1}{2}(P_{(0)}(V)+P_{(1)}(V)),\qquad P'(V)=\frac{1}{2}(P_{(0)}(V)-P_{(1)}(V)).\nonumber
\end{eqnarray}

\begin{thm}\label{theoremObSvEvOdd1} We have
\begin{eqnarray}
&&\beta_B H_{(0)}(U)=H_{(0)}(U) \gamma_B,\quad \gamma_B P_{(0)}(V)=P_{(0)}(V) \beta_B,\quad \forall B\in\I_{(0)},\nonumber\\
&&\beta_B H_{(0)}(U)=H_{(1)}(U) \gamma_B,\quad \gamma_B P_{(0)}(V)=P_{(1)}(V) \beta_B,\quad \forall B\in\I_{(1)},\nonumber\\
&&\beta_B H_{(1)}(U)=H_{(1)}(U) \gamma_B,\quad \gamma_B P_{(1)}(V)=P_{(1)}(V) \beta_B,\quad   \forall B\in\I_{(0)},\nonumber\\
&&\beta_B H_{(1)}(U)=H_{(0)}(U) \gamma_B,\quad \gamma_B P_{(1)}(V)=P_{(0)}(V) \beta_B,\quad  \forall B\in\I_{(1)},\nonumber
\end{eqnarray}
and
\begin{eqnarray}
&&H_{(0)}^2(U)=H_{(1)}^2(U)=H_{(0)}(U),\quad P_{(0)}^2(V)=P_{(1)}^2(V)=P_{(0)}(V),\nonumber\\ &&H_{(1)}(H_{(0)}(U))=H_{(0)}(H_{(1)}(U))=H_{(1)}(U),\nonumber\\
&&P_{(1)}(P_{(0)}(V))=P_{(0)}(P_{(1)}(V))=P_{(1)}(V),\nonumber\\
&&H_{(0)}(H(U))=H_{(1)}(H(U))=H(H_{(0)}(U))=H(H_{(1)}(U))=H(U),\nonumber\\
&&P_{(0)}(P(V))=P_{(1)}(P(V))=P(P_{(0)}(V))=P(P_{(1)}(V))=P(V).\nonumber
\end{eqnarray}
\end{thm}

\begin{proof} The proof is similar to the proof of Theorem \ref{theoremObSv1}.   \end{proof}

Using the previous theorems, we get the following multiplication tables (see Table \ref{tab1} and Table \ref{tab2}) for the operators $H$, $H'$, $H_{(0)}$, $H_{(1)}$ and the operators $P$, $P'$, $P_{(0)}$, $P_{(1)}$.

\begin{table}[ht]
\caption{Multiplication table for the operators $H$, $H'$, $H_{(0)}$, $H_{(1)}$.\label{tab1}}
\begin{tabular}{|c|c|c|c|c|}
\hline
  & $H$ & $H_{(0)}$ & $H_{(1)}$ & $H'$ \\ \hline
$H$ & $H$ & $H$ & $H$ & $0$ \\ \hline
$H_{(0)}$ & $H$ & $H_{(0)}$ & $H_{(1)}$ & $H'$ \\ \hline
$H_{(1)}$ & $H$ & $H_{(1)}$ & $H_{(0)}$ & $-H'$ \\ \hline
$H'$ & $0$ & $H'$ & $-H'$ & $H'$ \\ \hline
\end{tabular}
\end{table}

\begin{table}[ht]
\caption{Multiplication table for the operators $P$, $P'$, $P_{(0)}$, $P_{(1)}$.\label{tab2}}
\begin{tabular}{|c|c|c|c|c|}
\hline
  & $P$ & $P_{(0)}$ & $P_{(1)}$ & $P'$ \\ \hline
$P$ & $P$ & $P$ & $P$ & $0$ \\ \hline
$P_{(0)}$ & $P$ & $P_{(0)}$ & $P_{(1)}$ & $P'$ \\ \hline
$P_{(1)}$ & $P$ & $P_{(1)}$ & $P_{(0)}$ & $-P'$ \\ \hline
$P'$ & $0$ & $P'$ & $-P'$ & $P'$ \\ \hline
\end{tabular}
\end{table}

% \begin{table}[H] 
% \caption{This is a table caption. Tables should be placed in the main text near to the first time they are~cited.}
% \newcolumntype{C}{>{\centering\arraybackslash}X}
% \begin{tabularx}{\textwidth}{CCC}
% \toprule
% \textbf{Title 1}	& \textbf{Title 2}	& \textbf{Title 3}\\
% \midrule
% Entry 1		& Data			& Data\\
% Entry 2		& Data			& Data \textsuperscript{1}\\
% \bottomrule
% \end{tabularx}
% \noindent{\footnotesize{\textsuperscript{1} Tables may have a footer.}}
% \end{table}

\section{Relation between operators in the case of odd $n$}\label{sect5}

In this section, we prove that the operators considered in the previous section are related to the generalized Reynolds operators in the case of odd $n$.

We need the concept of adjoint multi-indices introduced in the previous work \cite{aver} of the author. We call ordered multi-indices $a_1 \ldots a_k$ and $b_1 \ldots b_l$ {\it adjoint multi-indices} if they have no common indices and they form multi-index $1\ldots n$ of the length $n$. We write $b_1\ldots b_l=\widetilde{a_1 \ldots a_k}$ and $a_1 \ldots a_k=\widetilde{b_1 \ldots b_l}$. We denote the sets of corresponding $2^{n-1}$ multi-indices by $\I_{\Adj}$ and $\widetilde{\I_{\Adj}}=\I \setminus \I_{\Adj}$. Therefore for each multi-index in $\I_{\Adj}$, there exists an adjoint multi-index in $\widetilde{\I_{\Adj}}$. We have $\mathfrak{B}=\{e_A \,|\, A\in\I \}=\{e_A \,|\, A\in\I_{\Adj}\} \cup \{e_A \,|\, A\in \widetilde{\I_{\Adj}}\}$. For example, $\I_{\Adj}=\I_{\First}$, $\widetilde{\I_{\Adj}}=\I\setminus\I_{\First}=\I_{\Last}$, where $\I_{\First}$ consists of the first (in the order) $2^{n-1}$ multi-indices of the set $\I$. In the case of odd $n$, we can write
$\I_{\First}=\{A\in\I,\, |A|\leq\frac{n-1}{2}\}$, $\I_{\Last}=\{A\in\I,\, |A|\geq\frac{n+1}{2}\}$. In the case of odd $n$, we can consider the following adjoint sets
$\I_{\Adj}=\I_{(0)}$, $\widetilde{\I_{\Adj}}=\I_{(1)}$.

Let us consider the following operators in $\cl^\F_{p,q}$ with odd $n=p+q$ for the sets (\ref{gamma}), (\ref{beta}), and some $\I_{\Adj}$, $\widetilde{\I_{\Adj}}$:
\begin{eqnarray}
&&H_{\Adj}(U)=\frac{1}{2^{n-1}}\!\!\sum_{A\in\I_{\Adj}}\!\!\beta_A U \gamma^A,\quad H_{\widetilde{\Adj}}(U)=\frac{1}{2^{n-1}}\!\!\sum_{A\in\widetilde{\I_{\Adj}}}\!\!\beta_A U \gamma^A,\nonumber\\
&&H(U)=\frac{1}{2}(H_{\Adj}(U)+H_{\widetilde{\Adj}}(U)),\qquad U\in\cl^\F_{p,q}.\nonumber
\end{eqnarray}

Using Lemma \ref{theoremSecBasReal}, we conclude that the elements $\gamma_{1\ldots n}$ and $\beta_{1\ldots n}$ are equal to  the elements $\pm e$ and $\pm e_{1\ldots n}$ in the case of odd $n$ and $\F=\R$, they are equal to the elements $\pm e$, $\pm ie$, and $\pm e_{1\ldots n}$ in the case of odd $n$ and $\F=\C$. Thus we have four (six in the complex case) different cases
$$\gamma_{1\ldots n}=\pm\beta_{1\ldots n},\qquad \gamma_{1\ldots n}=\pm e_{1\ldots n}\beta_{1\ldots n},\qquad \gamma_{1\ldots n}=\pm i e_{1\ldots n}\beta_{1\ldots n}.$$

We use notation
$$H_{\First}(U)=\frac{1}{2^{n-1}}\sum_{A\in\I_{\First}}\beta_A U \gamma^A,\qquad H_{\Last}(U)=\frac{1}{2^{n-1}}\sum_{A\in\I_{\Last}}\beta_A U \gamma^A$$
in the following theorem.

\begin{thm}\label{theoremObSvDual} Let us consider $\cl^\F_{p,q}$ with odd $n=p+q$ and the sets (\ref{gamma}), (\ref{beta}).
\begin{enumerate}
  \item If $\gamma_{1\ldots n}=\beta_{1\ldots n}$, then
  \begin{eqnarray}
H_{\Adj}(U)=H(U).\nonumber
\end{eqnarray}
In particular, we have
\begin{eqnarray}
 H(U)=H_{(0)}(U)=H_{(1)}(U)=H_{\First}(U)=H_{\Last}(U),\qquad H'(U)=0.\nonumber
\end{eqnarray}
  \item If $\gamma_{1\ldots n}=-\beta_{1\ldots n}$, then
    \begin{eqnarray}
H(U)=0,\qquad H_{\Adj}(U)=-H_{\widetilde{\Adj}}(U).\nonumber
\end{eqnarray}
In particular, we have
\begin{eqnarray}
H_{(0)}(U)=-H_{(1)}(U)=H'(U),\qquad H_{\First}(U)=-H_{\Last}(U).\nonumber
\end{eqnarray}
  \item If $\gamma_{1\ldots n}=e_{1\ldots n}\beta_{1\ldots n}$ (in the case $p-q\equiv 1 \mod4$), then
  \begin{eqnarray}
H_{\Adj}(U)=e_{1\ldots n}H_{\widetilde{\Adj}}(U),\qquad H(U)= \frac{e+e_{1\ldots n}}{2}H_{\Adj}(U).\nonumber
\end{eqnarray}
  \item If $\gamma_{1\ldots n}=-e_{1\ldots n}\beta_{1\ldots n}$ (in the case $p-q\equiv 1 \mod4$), then
\begin{eqnarray}
H_{\Adj}(U)=-e_{1\ldots n}H_{\widetilde{\Adj}}(U),\qquad H(U)= \frac{e-e_{1\ldots n}}{2}H_{\Adj}(U).\nonumber
\end{eqnarray}
  \item If $\gamma_{1\ldots n}=ie_{1\ldots n}\beta_{1\ldots n}$ (in the case $\F=\C$ and $p-q\equiv 3 \mod4$), then
  \begin{eqnarray}
H_{\Adj}(U)=ie_{1\ldots n}H_{\widetilde{\Adj}}(U),\qquad H(U)= \frac{e+ie_{1\ldots n}}{2}H_{\Adj}(U).\nonumber
\end{eqnarray}
  \item If $\gamma_{1\ldots n}=-ie_{1\ldots n}\beta_{1\ldots n}$ (in the case $\F=\C$ and $p-q\equiv 3 \mod4$), then
  \begin{eqnarray}
H_{\Adj}(U)=-ie_{1\ldots n}H_{\widetilde{\Adj}}(U),\qquad H(U)= \frac{e-ie_{1\ldots n}}{2}H_{\Adj}(U).\nonumber
\end{eqnarray}
In particular, we have in Cases 3 - 6:
\begin{eqnarray}
 H_{\Adj}(U)=H_{(0)}(U)=H_{(1)}(U)=H_{\First}(U)=H_{\Last}(U).\nonumber
\end{eqnarray}
\end{enumerate}

\end{thm}

\begin{proof} Using Lemma \ref{theoremSecBasReal}, we conclude that the elements $\beta_{1\ldots n}$ and $\gamma_{1\ldots n}$ are in the center of $\cl^\F_{p,q}$ with odd $n=p+q$. Thus, these elements commute with all elements of $\cl^\F_{p,q}$.

If $\beta_{1\ldots n}=\gamma_{1\ldots n}$, then
$$\beta_{a_1 \ldots a_m} U \gamma^{a_1 \ldots a_m}=\beta_{1\ldots n}\gamma^{1\ldots n}\beta_{a_1 \ldots a_m} U \gamma^{a_1 \ldots a_m}=\beta_{\widetilde{a_1\ldots a_m}}U\gamma^{\widetilde{a_1\ldots a_m}}.$$
Thus we obtain $\sum_A \beta_A U \gamma^A= 2 \sum_{A\in\I_{\Adj}}\beta_A F \gamma^A$ and the first statement of the theorem.

If $\beta_{1\ldots n}=-\gamma_{1\ldots n}$, then analogously
$\beta_{a_1 \ldots a_m} \!U \gamma^{a_1 \ldots a_m}\!=\!-\beta_{\widetilde{a_1\ldots a_m}}\!\!U\gamma^{\widetilde{a_1\ldots a_m}}\!$ and $$\sum_A\beta_A U \gamma^A=\sum_{A\in\I_{\Adj}} \beta_A U \gamma^A + \sum_{A\in\widetilde{\I_{\Adj}}} \beta_A U \gamma^A=0.$$

If $\gamma_{1\ldots n}=e_{1\ldots n}\beta_{1\ldots n}$, then
$$\beta_{a_1 \ldots a_m} U \gamma^{a_1 \ldots a_m}\!=\!\beta_{1\ldots n}e_{1\ldots n}\gamma^{1\ldots n}\beta_{a_1 \ldots a_m} U \gamma^{a_1 \ldots a_m}=e_{1\ldots n}\beta_{\widetilde{a_1\ldots a_m}}U\gamma^{\widetilde{a_1\ldots a_m}}.$$
Thus we obtain $\sum_A\beta_A U \gamma^A= (e+e_{1\ldots n}) \sum_{A\in\I_{\Adj}}\beta_A U \gamma^A$ and the third statement of the theorem.

If $\gamma_{1\ldots n}=-e_{1\ldots n}\beta_{1\ldots n}$, then
$\beta_{a_1 \ldots a_m} U \gamma^{a_1 \ldots a_m}\!=\!-e_{1\ldots n}\beta_{\widetilde{a_1\ldots a_m}}U\gamma^{\widetilde{a_1\ldots a_m}}$ and $$\sum_A\beta_A U \gamma^A= (e-e_{1\ldots n}) \sum_{A\in\I_{\Adj}}\beta_A U \gamma^A.$$

If $\gamma_{1\ldots n}=ie_{1\ldots n}\beta_{1\ldots n}$, then
$$\beta_{a_1 \ldots a_m}\! U \gamma^{a_1 \ldots a_m}\!=\!\beta_{1\ldots n}ie_{1\ldots n}\gamma^{1\ldots n}\beta_{a_1 \ldots a_m}\! U \gamma^{a_1 \ldots a_m}\!=\!ie_{1\ldots n}\beta_{\widetilde{a_1\ldots a_m}}\!U\gamma^{\widetilde{a_1\ldots a_m}}.$$
Thus we obtain $\sum_A\beta_A U \gamma^A= (e+ie_{1\ldots n}) \sum_{A\in\I_{\Adj}}\beta_A U \gamma^A$.

If $\gamma_{1\ldots n}=-ie_{1\ldots n}\beta_{1\ldots n}$, then
$\beta_{a_1 \ldots a_m} \!U \gamma^{a_1 \ldots a_m}\!=\!-ie_{1\ldots n}\beta_{\widetilde{a_1\ldots a_m}}\!U\gamma^{\widetilde{a_1\ldots a_m}}$ and $$\sum_A\beta_A U \gamma^A= (e-ie_{1\ldots n}) \sum_{A\in\I_{\Adj}}\beta_A U \gamma^A.$$
This completes the proof.  \end{proof}

We use Theorem \ref{theoremObSvDual} in the next section.

\section{Using generalized Reynolds operators to prove Pauli's theorem in Clifford algebras}
\label{sect6}

In this section, we show the application of generalized Reynolds operators to prove generalization of Pauli's theorem \cite{Pauli} to the case of Clifford algebras of arbitrary dimension. We use the theorems from the previous sections of this paper.

Note the following well-known fact. The Clifford algebras $\cl^\F_{p,q}$ with even $n=p+q$ and $\cl_{p,q}$, $p-q\equiv 3\mod 4$ are simple. The Clifford algebras $\cl_{p,q}$, $p-q\equiv 1\mod 4$ and $\C\otimes\cl_{p,q}$ with odd $n=p+q$ are not simple; they are direct sums of two simple algebras. The existence (or not existence) of an element $T$ that connect two different sets of Clifford algebra elements that satisfy the main anticommutative conditions can be proved using representation theory. The main idea of the following statements is to present an algorithm to compute this element $T$. These statements have been formulated in a brief report \cite{1} of the author without proof. In this paper, we demonstrate how we can prove them using generalized Reynolds operators.

We repeat the formulation of the theorems here (Theorems \ref{theoremPauliEv} and \ref{theoremPauliOddReal}) for consistency of the presentation. The proof of the theorems is new and show us the application of the method of averaging and generalized Reynolds operators.

\begin{thm} \cite{1} \label{theoremPauliEv} Let us consider $\cl^\F_{p,q}$ with even $n=p+q$. Let two sets of Clifford algebra elements
$\gamma_a, \beta_a, a=1, 2, \ldots, n$ satisfy conditions
\begin{eqnarray}
\gamma_a \gamma_b + \gamma_b \gamma_a= 2 \eta_{ab} e,\qquad \beta_a \beta_b + \beta_b \beta_a= 2 \eta_{ab} e.\nonumber
\end{eqnarray}
Then the sets $\{\gamma_A, A\in\I\}$ and $\{\beta_A, A\in\I\}$ are linearly independent (in the real case $\F=\R$, they are new bases of $\cl_{p,q}$) and there exists a unique (up to multiplication by a real (respectively, complex) number) element $T\in\cl^\F_{p,q}$ such that
\begin{eqnarray}
\gamma_{a}=T^{-1}\beta_a T,\qquad \forall a=1, \ldots, n. \nonumber
\end{eqnarray}
Additionally, we can obtain this element $T$ in the following way:
\begin{eqnarray}
T=H(U)=\frac{1}{2^n}\sum_A\beta_A U \gamma^A,\label{TT}
\end{eqnarray}
where $U$ is an element
\begin{itemize}
  \item of the set $\{ \gamma_A, A\in\I_{(0)}\}$ if $\beta_{1\ldots n}\neq-\gamma_{1\ldots n}$,
  \item of the set $\{ \gamma_A, A\in\I_{(1)}\}$ if $\beta_{1\ldots n}\neq\gamma_{1\ldots n}$,
\end{itemize}
such that $H(U)\neq 0$.
\end{thm}

\begin{proof}\!\!\! To obtain this statement, we use the properties of generalized Reynolds operators, namely, Theorems \ref{theoremObSv1}, \ref{theoremObSv3}, \ref{theoremObSv6}. Linearly independence of the sets follows from Lemma \ref{theoremSecBasReal}. For two arbitrary elements $U, V\in\cl^\F_{p,q}$ and elements (\ref{Q}), we have (\ref{QT}) by Theorem \ref{theoremObSv3}. There exists $U$ such that $H(U)$ is nonzero by Theorem \ref{theoremObSv6}. Further we take element $V$ such that $\pi_0(V H(U))\neq 0$ (we can take $V$ from the set of basis elements $\{e_A, A\in \I\}$). Therefore from (\ref{QT}), we see that $T=H(U)$ is invertible. Using Theorem \ref{theoremObSv1}, we obtain $\gamma_{a}=T^{-1}\beta_a T$, $\forall a=1, \ldots, n$. We obtain algorithm to compute the element $T$ (\ref{TT}) from Theorem~\ref{theoremObSv6}.

Let us prove that $T$ is unique up to multiplication by a constant. Suppose that we have two elements $T_1$, $T_2$ that satisfy $\gamma_{a}=T^{-1}\beta_a T$, $\forall a=1, \ldots, n$. Then for any $a=1, \ldots, n$, we have $T_1^{-1} \beta_a T_1=T_2^{-1} \beta_a T_2.$ Let us multiply both sides of this equation on the left by $T_1$ and on the right by $(T_2)^{-1}$. We obtain $[T_1 T_2^{-1}, \beta_a]=0$ for $a=1, \ldots, n$. Using
${\rm Cen}(\cl^\F_{p,q})=\cl^{0\,\F}_{p,q}$ ($n$ is even), we obtain $T_1=\mu T_2$, where $\mu\neq 0$, $\mu\in\F$. \end{proof}

To prove Theorem \ref{theoremPauliOddReal}, we need Lemmas \ref{theoremSecBasTrace} and \ref{theoremSecBas1}.

\begin{lem}\label{theoremSecBasTrace} Let us consider $\cl^\F_{p,q}$, $n=p+q$ and the set (\ref{oprs}).
\begin{enumerate}
  \item If $n=p+q$ is even, then $\pi_0(\gamma_{a_1\ldots a_k})=0,\quad k=1, \ldots, n.$
  \item If $n=p+q$ is odd, then $\pi_0(\gamma_{a_1\ldots a_k})=0,\quad k=1, \ldots, n-1$ and
  \begin{eqnarray}
\pi_0(\gamma_{1\ldots n})=\left\lbrace
\begin{array}{ll}
0, & \mbox{if (\ref{beta2}) is linearly independent};\\
\pm 1, \pm i, & \mbox{if (\ref{beta2}) is linearly dependent.}
\end{array}
\right.\nonumber
\end{eqnarray}
\end{enumerate}
Values $\pm i$ are possible only in the case of the complexified Clifford algebra.

If $n$ is odd, then
$\pi_n(\gamma_{a_1\ldots a_k})=0,\quad k=1, \ldots, n-1$ and
\begin{eqnarray}
\pi_n(\gamma_{1\ldots n})=\left\lbrace
\begin{array}{ll}
\pm 1, & \mbox{if (\ref{beta2}) is linearly independent};\\
0, & \mbox{if (\ref{beta2}) is linearly dependent.}
\end{array}
\right.\nonumber
\end{eqnarray}
\end{lem}

\begin{proof} For any element $\gamma_A$ (except $\gamma_{1\ldots n}$ in the case of odd $n$ and except $e$ in the case of any $n$), there exists element $\gamma_a$ such that $\gamma_A$ anticommutes with $\gamma_a$ (if $|A|$ is even, then we can take $a\in A$; if $|A|$ is odd, then we can take $a\notin A$).
We obtain
$$\pi_0(\gamma_A)=\pi_0(-\gamma_a\gamma_A(\gamma_a)^{-1})=-\pi_0(\gamma_A)$$
and $\pi_0(\gamma_A)=0.$  Further, we use Lemma \ref{theoremSecBasReal}.

The statement for the operator $\pi_n$ can be proved similarly using the following property: for any two elements $U, V$ of $\cl^\F_{p,q}$ with odd $n=p+q$, we have $\pi_n(UV)=\pi_n(VU).$ It follows from the fact $\pi_n([U,V])=0$ (see \cite{Sh1}). \end{proof}

\begin{lem}\label{theoremSecBas1}
\begin{enumerate}
  \item Let us consider $\cl^\F_{p,q}$ with odd $n=p+q$ such that $p-q\equiv 1 \mod 4$, the set (\ref{oprs}), and elements
\begin{eqnarray}
  \sigma_a=e_{1\ldots n}\gamma_{a},\qquad a=1, \ldots, n\nonumber
\end{eqnarray}
  \item Let us consider $\C\otimes\cl_{p,q}$ with odd $n=p+q$ such that $p-q\equiv 3 \mod 4$, the set (\ref{oprs}) and elements
\begin{eqnarray}
  \sigma_a=i e_{1\ldots n}\gamma_a,\qquad a=1, \ldots, n\nonumber
\end{eqnarray}
\end{enumerate}
In both cases the elements $\sigma_a$, $a=1, \ldots, n$ satisfy conditions $\sigma_a \sigma_b + \sigma_b \sigma_a= 2 \eta_{ab} e.$ If $\{\gamma_A, A\in\I\}$ is linearly independent, then $\{\sigma_A, A\in\I\}$ is linearly dependent. If $\{\gamma_A, A\in\I\}$ is linearly dependent, then $\{\sigma_A, A\in\I\}$ is linearly independent.
\end{lem}

\begin{proof} (1) Using (\ref{oprs}), we get
$$\sigma_a \sigma_b + \sigma_b \sigma_a=e_{1\ldots n}\gamma_a e_{1\ldots n}\gamma_b + e_{1\ldots n}\gamma_b e_{1\ldots n}\gamma_a =2 \eta_{ab} e,$$
because $e_{1\ldots n}\in {\rm Cen}(\cl^\F_{p,q})$ and $(e_{1\ldots n})^2=(-1)^{\frac{n(n-1)}{2}}(-1)^q e=e$ because $p-q\equiv 1 \mod 4$.

If $\gamma_{1\ldots n}=\pm e_{1\ldots n}$, then $\sigma_{1\ldots n}=\pm(e_{1\ldots n})^n \gamma_{1\ldots n}=\pm e_{1\ldots n}e_{1\ldots n}=\pm e$. If $\gamma_{1\ldots n}=\pm e$, then $\sigma_{1\ldots n}=\pm(e_{1\ldots n})^n \gamma_{1\ldots n}=\pm e_{1\ldots n}e=\pm e_{1\ldots n}$. Further we use Lemma \ref{theoremSecBasReal}.

(2) The proof of the second statement of the lemma is similar to the proof of the first statement. We have
$$\sigma_a \sigma_b + \sigma_b \sigma_a=i e_{1\ldots n}\gamma_a i e_{1\ldots n}\gamma_b + i e_{1\ldots n}\gamma_b i e_{1\ldots n}\gamma_a =2 \eta_{ab} e,$$ because $e_{1\ldots n}\in {\rm Cen}(\C\otimes\cl_{p,q})$ and $(e_{1\ldots n})^2=(-1)^{\frac{n(n-1)}{2}}(-1)^q e=-e$ because $p-q\equiv 3 \mod 4$.

If $\gamma_{1\ldots n}=\pm e_{1\ldots n}$, then $\sigma_{1\ldots n}=\pm(i e_{1\ldots n})^n \gamma_{1\ldots n}=\pm i e_{1\ldots n}e_{1\ldots n}=\pm i e$. If $\gamma_{1\ldots n}=\pm i e$, then $\sigma_{1\ldots n}=\pm(ie_{1\ldots n})^n \gamma_{1\ldots n}=\pm ie_{1\ldots n}ie=\pm e_{1\ldots n}$.  \end{proof}

\begin{thm} \cite{1} \label{theoremPauliOddReal} Let us consider $\cl^\F_{p,q}$ with odd $n=p+q$. Suppose that two sets of Clifford algebra elements $\gamma_a, \beta_a, a=1, 2, \ldots, n$ satisfy the conditions
\begin{eqnarray}
\gamma_a \gamma_b + \gamma_b \gamma_a= 2 \eta_{ab} e,\qquad \beta_a \beta_b + \beta_b \beta_a= 2 \eta_{ab} e. \nonumber
\end{eqnarray}
Then, for $\cl^\F_{p,q}$ of signature $p-q\equiv 1 \mod4$, the element $\gamma_{1\ldots n}$ either takes the values $\pm e_{1\ldots n}$ and the set $\{\gamma_A, A\in\I\}$ is linearly independent or takes the values $\pm e$ and then the set is linearly dependent. The same is for the set $\{\beta_A, A\in\I\}$. We have cases 1--4 below.

For $\cl^\F_{p,q}$ of signature $p-q\equiv 3 \mod4$, the element $\gamma_{1\ldots n}$ either takes the values $\pm e_{1\ldots n}$ and the set $\{\gamma_A, A\in\I\}$ is linearly independent or takes the values $\pm ie$ (this is possible only in the case $\F=\C$) and then the set is linearly dependent. The same is for the set $\{\beta_A, A\in\I\}$. We have cases 1--2, 5--6 below.

There exists a unique element $T$ of $\cl^\F_{p,q}$ (up to multiplication by an invertible element of the center of $\cl^\F_{p,q}$) such that
\begin{eqnarray}
&1.& \gamma_{a}=T^{-1}\beta_a T,\quad  \forall a=1, \ldots, n  \quad\Leftrightarrow\, \beta_{1\ldots n}=\gamma_{1\ldots n}\nonumber;\\
&2.& \gamma_{a}=-T^{-1}\beta_a T,\quad  \forall a=1, \ldots, n  \quad\Leftrightarrow\, \beta_{1\ldots n}=-\gamma_{1\ldots n} \nonumber;\\
&3.& \gamma_{a}=e_{1\ldots n}T^{-1}\beta_a T,\quad  \forall a=1, \ldots, n  \quad\Leftrightarrow\, \beta_{1\ldots n}=e_{1\ldots n}\gamma_{1\ldots n}\nonumber;\\
&4.& \gamma_{a}=-e_{1\ldots n}T^{-1}\beta_a T,\quad  \forall a=1, \ldots, n  \quad\Leftrightarrow\, \beta_{1\ldots n}=-e_{1\ldots n}\gamma_{1\ldots n} \nonumber;\\
&5.& \gamma_{a}=ie_{1\ldots n}T^{-1}\beta_a T,\quad  \forall a=1, \ldots, n  \quad\Leftrightarrow\, \beta_{1\ldots n}=ie_{1\ldots n}\gamma_{1\ldots n}\nonumber;\\
&6.& \gamma_{a}=-ie_{1\ldots n}T^{-1}\beta_a T,\quad  \forall a=1, \ldots, n  \quad\Leftrightarrow\, \beta_{1\ldots n}=-ie_{1\ldots n}\gamma_{1\ldots n} \nonumber.
\end{eqnarray}

For all cases, we have $\gamma_a=(\beta_{1\ldots n}\gamma^{1\ldots n})T^{-1}\gamma_a T$, $a=1, \ldots, n$.

Additionally, in the case of the real Clifford algebra $\cl_{p,q}$ of signature $p-q\equiv 1 \mod 4$ and the complexified Clifford algebra $\C\otimes\cl_{p,q}$ of arbitrary signature, the element $T$, whose existence is stated in cases 1--6 of the theorem, equals
\begin{eqnarray}T=H_{(0)}(U)=\frac{1}{2^{n-1}}\sum_{A\in\I_{(0)}}\beta_A U \gamma_A^{-1}\label{TTT}\end{eqnarray}
where $U$ is an element of the set $\{ \gamma_A+\gamma_B,\, A, B\in\I_{(0)}\}.$

In the case of the real Clifford algebra $\cl_{p,q}$ of signature $p-q\equiv 3 \mod4$, the element $T$, whose existence is stated in cases 1 and 2 of the theorem, equals $T=H_{(0)}(U)$, where $U$ is element of the set $\{\gamma_A,\, A\in\I_{(0)}\}$ such that $H_{(0)}(U)\neq 0$.
\end{thm}

\begin{proof} Linearly independence (or linearly dependence) of the sets follows from Lemma \ref{theoremSecBasReal}. This implies that we have four cases in $\cl_{p,q}$ and six cases in $\C\otimes\cl_{p,q}$.

Cases 3, 4, 5 and 6 of the theorem are reduced to cases 1 and 2 by Lemma \ref{theoremSecBas1}. We must change one of the given sets by the set $\sigma_a$ in these cases. Case 2 of the theorem (when we have $\beta_{1\ldots n}=-\gamma_{1\ldots n}$) is reduced to case 1. We must consider the set
$\sigma_a=-\beta_a$ for $a=1, \ldots, n$. For this set, we have $\sigma_{1\ldots n}=(-1)^n \beta_{1\ldots n}=-\beta_{1\ldots n}=\gamma_{1\ldots n}$ and obtain
$\gamma_{a}=T^{-1}\sigma_a T=-T^{-1}\beta_a T.$

Thus we will consider and prove only case 1 of the theorem (when $\beta_{1\ldots n}=\gamma_{1\ldots n}$). We will consider only the case $\beta_{1\ldots n}=\gamma_{1\ldots n}=\pm e_{1\ldots n}$ (other cases are reduced to this case by Lemma \ref{theoremSecBas1}).

Let us consider arbitrary elements $U, V\in\cl^\F_{p,q}$ and expressions (\ref{Q}). Then we have (\ref{obshcommut}) and (\ref{QTnech}) by Theorems \ref{theoremObSv1} and \ref{theoremObSv3}.
We must prove that there exist elements $H(U)$ and $V$ such that $\pi_0(VH(U))e+\pi_n(VH(U))e_{1\ldots n}$ is invertible. Then from (\ref{QTnech}), we will see that $T=H(U)$ is invertible, and from (\ref{obshcommut}), we will obtain $\gamma_{a}=T^{-1}\beta_a T$.
We have
\begin{eqnarray}
&&(\pi_0(V H(U))e+\pi_n(V H(U))e_{1\ldots n})(\pi_0(V H(U))e-\pi_n(V H(U))e_{1\ldots n})\nonumber\\
&&=(\pi_0^2(V H(U))-\pi_n^2(V H(U))(-1)^\frac{n(n-1)}{2}(-1)^q)e\\
&&=(\pi_0^2(V H(U))\pm\pi_n^2(V H(U)))e,\nonumber\end{eqnarray}
where sign ``$+$'' is in the case $p-q\equiv 3 \mod4$ and sign ``$-$'' is in the case $p-q\equiv 1 \mod4$.

I) Let us consider the case $p-q\equiv 1 \mod4$. We must choose elements $H(U)$ and $V$ such that $\pi_0(H(U) V)\neq\pm\pi_n(H(U) V).$
By Theorem \ref{theoremObSv6}, there exists element $U\in\{\gamma_B, B\in\I_{(0)} \}$ such that $H(U)\neq 0$. Since $\{\gamma_A, A\in\I\}$ is a basis, $H(U)$ can be written in the form $H(U)=\sum_A h_A\gamma_A$, $h_A\in\F$.

If there exists multi-index $C$ such that $h_C\neq 0$ and $h_{\widetilde C}\neq \pm h_C$ (where $\widetilde C$ and $C$ are adjoint multi-indices, see the previous section and \cite{aver}), then we can take $V=\gamma_C$. Using Lemma \ref{theoremSecBasTrace}, we obtain
$$\pi_0(H(U) V)=h_C\neq \pm h_{\widetilde{C}}=\pm \pi_n(H(U) V).$$
If there is no such index $C$, then $H(U)$ can be represented in the form
\begin{eqnarray}
H(U)=\sum_{j=1}^k \lambda_j \gamma_{A_j}(e+e_{1\ldots n})+\sum_{j=1}^l \mu_j \gamma_{B_j}(e-e_{1\ldots n}),\qquad \lambda_j, \mu_j\neq 0,\label{vid}
\end{eqnarray}
where all multi-indices $A_j$ and $B_j$ are different and any two of them do not constitute the full multi-index $1\ldots n$.

Using Theorem \ref{theoremObSv5}, we get $\sum_{B\in\I_{(0)}}H(\gamma_B)\gamma^B=e$. We have at least one $U\in\{\gamma_B, B\in\I_{(0)}\}$ such that $k\neq 0$ and at least one $U\in\{\gamma_B, B\in\I_{(0)}\}$ such that $l\neq 0$. Therefore there exists element $U\in\{\gamma_A+\gamma_B,\, A\neq B, A, B\in\I_{(0)}\}$ such that $k, l\neq 0$ in (\ref{vid}). Taking (\ref{vid}) and the element
\begin{eqnarray}V=\sum_{j=1}^k\frac{1}{\lambda_j} \gamma^{A_j}+\sum_{j=1}^l\frac{1}{\mu_j} \gamma^{B_j},\nonumber\end{eqnarray}
we obtain
\begin{eqnarray}
H(U)V=k(e+e_{1\ldots n})+(\sum_{i, j=1, i\neq j}^k \frac{\lambda_j}{\lambda_i}\gamma_{A_j}\gamma^{A_i} + \sum_{j=1}^k\sum_{i=1}^l\frac{\lambda_j}{\mu_i} \gamma_{A_j}\gamma^{B_i})(e+e_{1\ldots n})\nonumber\\
+l(e-e_{1\ldots n})+(\sum_{j=1}^l\sum_{i=1}^k \frac{\mu_j}{\lambda_i}\gamma_{B_j}\gamma^{A_i}+ \sum_{i, j=1, i\neq j}^l\frac{\mu_j}{\mu_i}\gamma_{B_j}\gamma^{B_i})(e-e_{1\ldots n}),\nonumber
\end{eqnarray}
and $k+l=\pi_0(H(U)V)\neq\pm\pi_n(H(U)V)=\pm(k-l).$

II) Let us consider the case $p-q\equiv 3 \mod4$. We must choose elements $H(U)$ and $V$ such that $\pi_0^2(V H(U))+\pi_n^2(V H(U))\neq 0$.
By Theorem \ref{theoremObSv6}, we always have element $U\in\{\gamma_A, A\in\I_{(0)}\}$ such that $H(U)\neq 0$. We can always take element $V\in\{e_A, A\in\I\}$ such that $\pi_0(V H(U))\neq 0$ or $\pi_n(V H(U))\neq 0$. In the case of the real Clifford algebra, the theorem is proved.

In the case of the complexified Clifford algebra, we must choose elements $H(U)$ and $V$ such that $\pi_0(H(U) V)\neq\pm i\pi_n(H(U) V).$  Further proof is similar to the proof of the case $p-q\equiv 1 \mod4$, but we consider the elements
\begin{eqnarray}
H(U)=\sum_{j=1}^k \lambda_j \gamma_{A_j}(e+ie_{1\ldots n})+\sum_{j=1}^l \mu_j \gamma_{B_j}(e-ie_{1\ldots n}),\qquad \lambda_j, \mu_j\neq 0\label{vid2}
\end{eqnarray}
instead of the elements (\ref{vid}).

Proof of uniqueness of element $T=H(U)$ up to multiplication by an invertible Clifford algebra element is similar to the proof of uniqueness in the case of even $n$.

According to the proof above, we can find element $T$ in different cases in the following form (up to multiplication by a nonzero constant):
\begin{enumerate}
  \item $\beta_{1\ldots n}=\gamma_{1\ldots n} \,\Rightarrow\, T=\sum_A\beta_A U \gamma^A,$
  \item $\beta_{1\ldots n}=-\gamma_{1\ldots n}\,\Rightarrow\,T=\sum_A(-1)^{|A|}\beta_A U \gamma^A,$
  \item $\beta_{1\ldots n}=e_{1\ldots n}\gamma_{1\ldots n}\,\Rightarrow\,T=\sum_{A\in\I_{(0)}}\beta_A U \gamma^A+e_{1\ldots n}\sum_{A\in\I_{(1)}}\beta_A U \gamma^A,$
  \item $\beta_{1\ldots n}=-e_{1\ldots n}\gamma_{1\ldots n}\,\Rightarrow\,T=\sum_{A\in\I_{(0)}}\beta_A U \gamma^A-e_{1\ldots n}\sum_{A\in\I_{(1)}}\beta_A U \gamma^A,$
  \item $\beta_{1\ldots n}=ie_{1\ldots n}\gamma_{1\ldots n}\,\Rightarrow\,T=\sum_{A\in\I_{(0)}}\beta_A U \gamma^A+ie_{1\ldots n}\sum_{A\in\I_{(1)}}\beta_A U \gamma^A,$
  \item $\beta_{1\ldots n}=-ie_{1\ldots n}\gamma_{1\ldots n}\,\Rightarrow\,T=\sum_{A\in\I_{(0)}}\beta_A U \gamma^A-ie_{1\ldots n}\sum_{A\in\I_{(1)}}\beta_A U \gamma^A.$
\end{enumerate}
Using Theorem \ref{theoremObSvDual}, we conclude that all these elements $T$ equal (\ref{TTT}) up to multiplication by a nonzero constant.
\end{proof}

%\section{Conclusion}\label{sect7}
\section{Discussion}\label{sect7}

In this paper, we develop the method of averaging in Clifford algebra suggested by the author in \cite{aver, averaging2}. We consider specific operators (generalized Reynolds operators) in Clifford algebras and study their properties (see Sections \ref{sect3}--\ref{sect5}). These operators allowed us to deduce an algorithm to compute elements that connect two different sets of Clifford algebra elements that satisfy the main anticommutative condition of Clifford algebra (see Section~\ref{sect6}).

The results obtained in this paper are used in geometry, physics, engineering, and other applications. We use the results in the study of $n$-dimensional Weyl, Majorana, and Majorana--Weyl spinors \cite{3} and in the theory of spin groups \cite{5, 4}. Using the algorithm from this paper, the method of calculating of elements of spin groups is presented. Some modification of this algorithm using zero divisors is discussed in \cite{KMC, KMC2} by other authors. The results of this paper are also used in problems related to the Dirac equation and spinors \cite{Das, Monakhov1, Monakhov2}, the Higgs model \cite{Maas}, applications of Riemannian geometry \cite{Gu}. Note other possible applications of the results in geometry, engineering, physics, and analysis \cite{Ma, Bisi, Adj, Bizi}.
%\cite{Das}, the Higgs model \cite{Maas}, applications of Riemannian geometry \cite{Gu}. 
We use generalized Reynolds operators in the proof of the local Pauli's theorem \cite{local}, when both sets of Clifford algebra elements depend on the point of Euclidean space. These operators can be used to study the same question and other problems for the Dirac and the Yang--Mills equations on non-trivial curved manifolds. The technique developed in this paper can be generalized to the case of other algebraic (for example, to matrix algebras using the Cartan--Bott 8-periodicity) and geometric structures (Atiyah--K\"ahler algebras \cite{IL,Atiyah,Kahler,AKS,OS} and the algebra of h-forms \cite{Primit,CC,hforms}, which are geometric generalizations of Clifford algebras).

\section*{Acknowledgments}

%The publication was prepared within the framework of the Academic Fund Program at the HSE University in 2022 (grant 22-00-001).

The author is grateful to the participants of a number of recent conferences (AGACSE, ICMMP, Alterman Conference), where the results of this paper were presented, for fruitful discussions. The author is grateful to three anonymous reviewers for their valuable comments on how to improve the paper.

This work is supported by the Russian Science Foundation (project 23-71-10028), https://rscf.ru/en/project/23-71-10028/.

\medskip

\noindent{\bf Data availability} Data sharing not applicable to this article as no datasets were generated or analyzed during the current study.

\medskip

\noindent{\bf Declarations}\\
\noindent{\bf Conflict of interest} The authors declare that they have no conflict of interest.

\bibliographystyle{spmpsci}

% % ------------------------------------------------------------------------
\end{document}